\documentclass[reqno]{amsart}
\usepackage{epsf}
\def\be{\begin{equation}}
\def\ee{\end{equation}}
\def\ba{\begin{array}}
\def\ea{\end{array}}
\def\bea{\begin{eqnarray}}
\def\eea{\end{eqnarray}}

\def\noi{\noindent}

\begin{document}

\vspace{-4truecm} %
{}\hfill{DSF$-$31/2007}%

\vspace{1truecm}

\title{Fermi at Los Alamos and the early Britain's way to nuclear
energy}

\author{S. Esposito}
\address{{\it S. Esposito}: Dipartimento di Scienze Fisiche,
Universit\`a di Napoli ``Federico II'' \& I.N.F.N. Sezione di
Napoli, Complesso Universitario di M. S. Angelo, Via Cinthia,
80126 Napoli ({\rm Salvatore.Esposito@na.infn.it})}%

\begin{abstract}
A novel recovery of some important documents related to the Fermi
course on neutron physics, held at Los Alamos in 1945, is
announced. Its relevance for the effective launch of a British
nuclear programme in the early postwar period, independently of
the U.S. technical cooperation (precluded by the American
authorities) and warmly supported by Chadwick, is discussed.
\end{abstract}

\maketitle



\noi The first achievements leading to the exploitation of the
enormous amount of the energy trapped inside the atom have been,
as for all scientific achievements, just part of the search of
science for a fuller explanation of Nature and the world around
us. Quite differently from other cases, however, the world
realized abruptly of the power of nuclear energy as early as
August 1945, with the well-known events that led to the end of the
second world war. Once the facts demonstrated that some correct
knowledge was reached on that part of Nature's functioning, time
came for the scientist, even for the ones who contributed to those
achievements, to pause and reason a bit more on the results
obtained.

As far as we know, the first occasion for this was the course on
neutron physics that Enrico Fermi taught at Los Alamos in the fall
of 1945 \cite{FNM}. The teacher and the students (about thirty) of
such a course, in fact, were some of the people who themselves
contributed, with different tasks, to the Manhattan Project or
other similar projects. Unfortunately enough, for some time only
the attending students benefitted from the teaching of Fermi,
ranging over more than ten years of important discoveries, since
the circulation of the notes taken down in class was limited by
presumed secrecy reasons.

The secrecy affair dates back to 1939, just after the discovery of
nuclear fission by Otto Hahn and Fritz Strassmann in Germany, and
to this regard the words of Fermi himself are illuminating:%
\begin{quote}
A world war was about to start. The new possibilities [opened by
the discovery of the fission of the uranium atom] appeared likely
to be important, not only for peace but also for war.

A group of physicists in the United States -- including Leo
Szilard, Walter Zinn, Herbert Anderson, and myself -- agreed
privately to delay further publications of the findings in this
field.

We were afraid these finding might help the Nazis.

Our action, of course, represented a break with the scientific
tradition and was not taken lightly. Subsequently, when the
government became interested in the atom bomb project, secrecy
became compulsory. [...]

Secrecy that we thought was an unwelcome necessity of the war
still appears to be an unwelcome necessity. \cite{CST}
\end{quote}

The ``unwelcome necessity'' was formally revealed by scientists of
the Chicago branch of the Manhattan Project as early as June 1945
in the {\it Franck Report}: ``Nuclear bombs cannot possibly remain
a `secret weapon' at the exclusive disposal of this country for
more than a few years. The scientific facts on which construction
is based are well known to scientists of other countries''
\cite{Cardozo}.

The expectations of these scientists, for removing the wartime
secrecy and restoring nuclear science to the realm of open
scientific inquiry, came together in the McMahon Bill of December
1945. The intent was, indeed, ``the free dissemination of basic
scientific information,'' ``as freely as may be consistent with
the foreign and domestic policies'' \cite{Cardozo} However the
final act, approved by the American Congress and signed by the
President on August 1946 as the Atomic Energy Act, changed the
original vision, and became ``a program for the control of
scientific and technical information.''

A Declassification Guide regarding, among the others, papers
dealing with the pile theory and the theory of neutron diffusion
was issued as early as 30 March 1946, and the subjects were
discussed in several meeting of the Basic Responsible Reviewers of
the Atomic Energy Commission (A.E.C.).\footnote{See a letter by
Richard T. Batson conserved among the Chadwick papers at the
Chur\-chill Archive Centre, Cambridge (U.K.), in the folder CHAD I
17/3; see below.} Some of the papers first released by the
Declassification Office, which served also as a guide for
declassification of further papers on the same subject, were the
following (see the mentioned letter by Batson): the first eight
chapter of {\it Neutron Physics}, the lecture course by Fermi
(notes written by I. Halpern), the papers {\it Elementary Theory
of the Pile} by E. Fermi, {\it Elementary Pile Theory} by S.K.
Allison and {\it Theoretical Discussion of a Small Homogeneous and
Enriched Detector} by R.F. Christy. Indeed, the scientific
community first knew about the detailed general functioning of an
atomic pile from the Fermi's paper mentioned just above, which was
published in {\it Science} in 1947 \cite{Science}.

\begin{figure}
\begin{center}
\vspace{1.5truecm} %
\epsfysize=17.6cm 
\epsffile{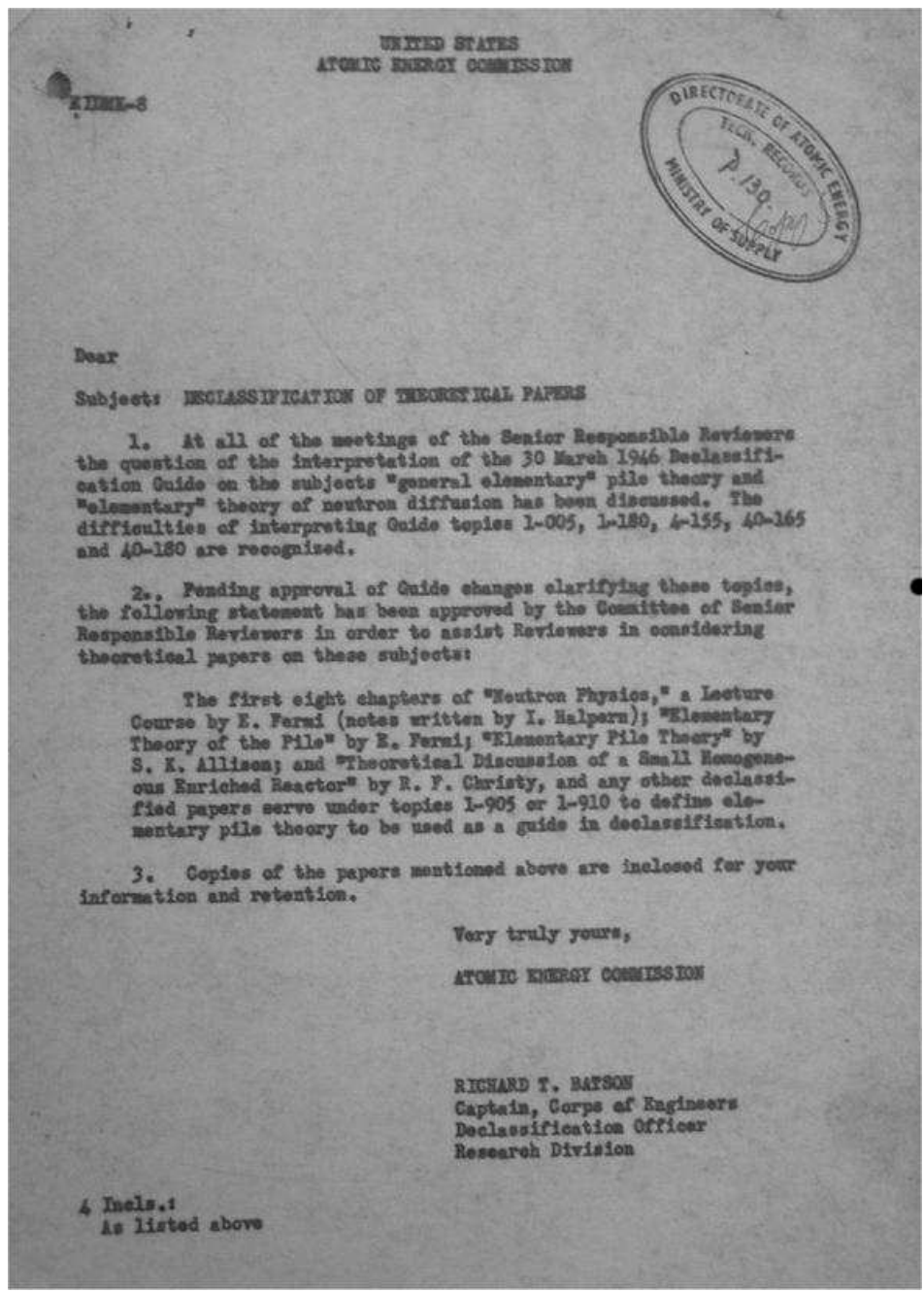}
\caption{The letter from Richard T. Batson conserved among the
Chadwick's papers, regarding the declassification of several
documents about nuclear studies. (Courtesy of the Churchill
Archive Centre, Cambridge).} \label{fig1}
\end{center}
\end{figure}

A first part of the Fermi lectures at Los Alamos, containing
neutron physics without reference to chain reactions, was in fact
declassified on 5 September 1946, while the remaining part has
been declassified only in 1962. Both parts have been later
published in the {\it Collected Papers} by Fermi \cite{FNM}.
Leaving aside the pregnant didactic style by Fermi, the main
relevance of such notes is that they present, for the first time,
a complete and accurate treatment of neutron physics from its
beginning, including a detailed study of the physics of the atomic
piles. In this respect it is not surprising that especially the
second part of the notes, dealing just with chain reactions and
pile physics, was considered as ``confidential'' material by
governmental offices.

Nevertheless, something happened that eluded the control network
for such information.

We have, in fact, recently recovered a {\it different} version of
the Fermi lectures at Los Alamos, formerly belonged to James
Chadwick and now deposited at the Chur\-chill Archive Centre in
Cambridge (U.K.). The folders relevant to us are essentially two.
The first one (CHAD I 17/3) contains a letter from the A.E.C.,
already mentioned above, a copy of the paper {\it Elementary
Theory of the Pile} by Fermi\footnote{This paper is reproduced in
Ref. \cite{FNM}; in particular see page 538 of the second volume.}
and a copy of only the {\it first part} of the Halpern notes of
the Fermi lectures. Instead the second folder (CHAD I 4/1)
contains a version of the {\it complete} set of lectures made by
A.P. French, dated June 23, 1947.

It is probably not strange\footnote{Note however that only a typed
copy of the letter from A.E.C., without the name of the addressee
and unsigned, is present in the folder, while it contains no copy
of the papers by Allison and Christy mentioned above (the letter
explicitly refers to them and to the Fermi papers).} that the
material of the first folder belonged to Chadwick, since he was
the respected (also by Americans) leader of the British Mission in
the United states.

The British Mission was formed after the Quebec Agreement on
Anglo-American collaboration was signed jointly by W. Chur\-chill
and F.D. Roosevelt. Almost all the work carried out in Great
Britain on $^{235}$U and atomic bomb calculations was suspended,
and a number of British scientists headed by Chadwick were
transferred to the United States. They joined to work in different
parts of the American programme for the nuclear energy, but the
biggest part of the British contingent was at Los Alamos, and
Chadwick himself was later present at the world's first nuclear
test at Alamogordo on 16 July 1945.

Several scientists of the British Mission were very young and,
among the others, it was Anthony P. French who graduated in
Physics at the Cambridge University just in 1942. In the same year
he joined the atomic bomb project (``Tube Alloys'') at the
Cavendish Laboratory, and was later sent to Los Alamos in October
1944 as a member of the British Mission. Here he worked with E.
Bretscher, O.R. Frisch, J. Hughes, D.G. Marshall, P.B. Moon, M.J.
Poole, J. Rotblat, E.W. Titterton and J.L. Tuck in the field of
experimental nuclear physics \cite{Szasz},\footnote{The remaining
part of the British Mission was composed by B. Davison, K. Fuchs,
D.J. Littler, W.G. Marley, R.E. Peierls, W.G. Penney, G. Placzek,
H. Sheard and T.H.R. Skyrmes.} and returned to the United Kingdom
in 1946, working for two years at the just newly formed Atomic
Energy Research Establishment (A.E.R.E.). The second folder of the
Chadwick papers mentioned above contains just the notes of Fermi
course on neutron physics taken by French on his own, when he was
at Los Alamos, and later (1947) re-organized into a final version
when he came back to England.

A preliminary study shows that the French notes do {\it not}
depend on the Halpern ones, but French probably saw them (the
organization of the introduction is similar). The topics covered
are exactly the same, although to a certain extent the material is
organized in a little different manner, but the text of the notes
is different in the two versions. In few cases, however, similar
or even identical words or sentences are present in both versions,
likely denoting quotes from an original wording by Fermi. In
general, the French notes are much more detailed (and accurate),
with a great number of shorter or larger peculiar
additions\footnote{Our preliminary study has shown that the case
is completely different, for example, from that of the revision of
the (first part of the) Halpern notes made by J.G. Beckerley in
1951 (document AECD 2664 of the Atomic Energy Commission). Here
the author {\it re-wrote} the Fermi lectures by including several
additions from {\it other} sources, ``where clarity demanded more
information and where the addition of recent data made the text
more complete.'' Contrarily to the present case (as it is evident
from the text of the notes), Beckerley ``was not privileged to
attend the course'' by Fermi.} (explanations, calculations, data
or other, and 5 more exercises) not present in the Halpern notes:
at least 15 of these additions are quite relevant. Instead the
peculiar additions present in the Halpern version but not in the
French one are very few (including 3 more exercises), only one of
them being relevant. Also, the French paper contains the six
questions which were set as a final examination at the end of the
lecture course.

\begin{figure}
\begin{center}
\vspace{1.5truecm} %
\epsfysize=17.6cm 
\epsffile{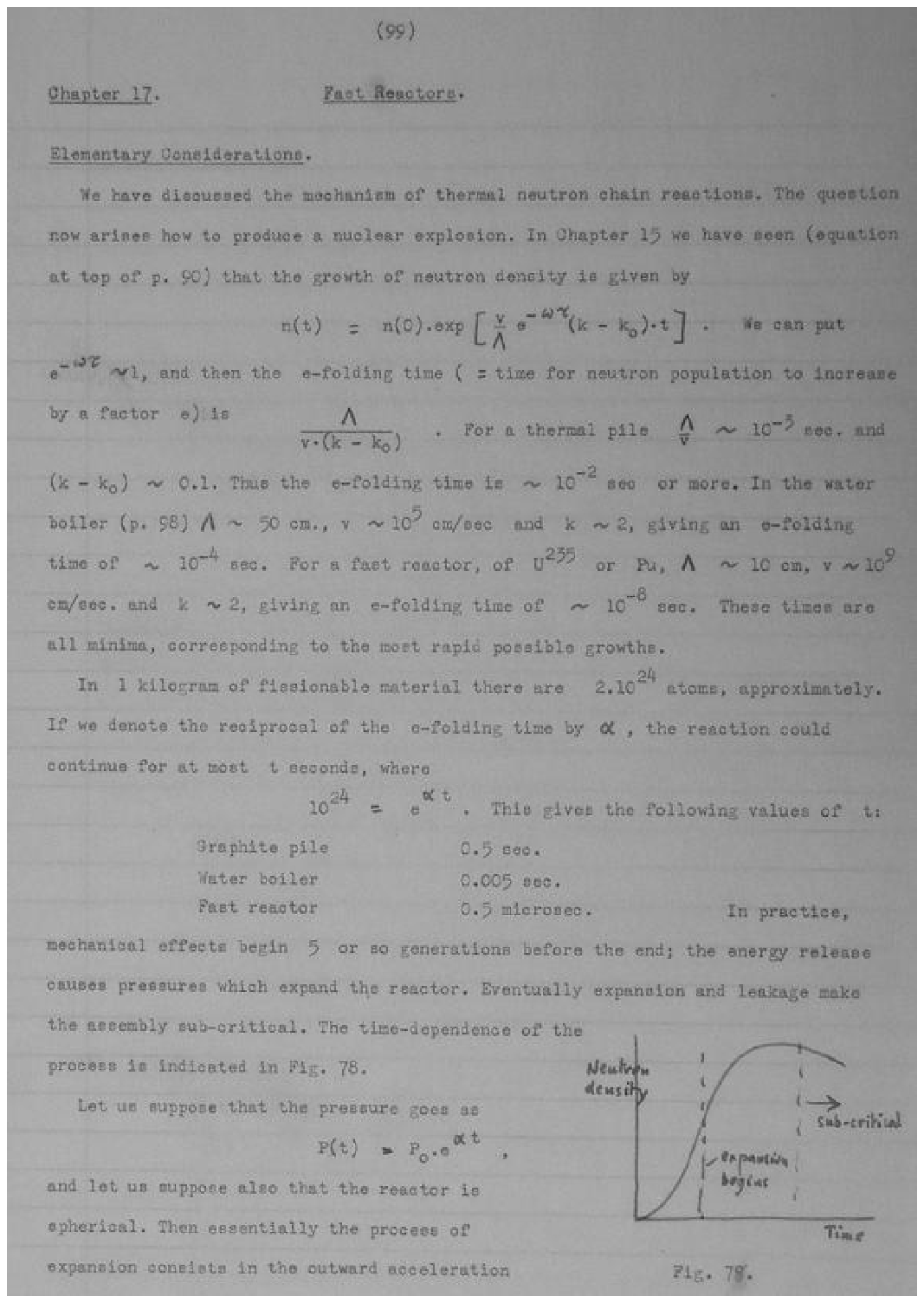}
\caption{One page excerpted from the French version of the Fermi
lecture notes on neutron physics. Note the explicit reference to a
nuclear explosion in the second line of the text. (Courtesy of the
Churchill Archive Centre, Cambridge).} \label{fig2}
\end{center}
\end{figure}

It is quite interesting that the greater detail already present in
the French notes increases even more in quality (especially
figures and data) in the last part, directly related to chain
reactions and their applications, and, moreover, explicit
references to bomb applications are made.

Finally, regarding Fermi's involvement in his own delivery of the
lectures, differently from Halpern, French limits the subjects
treated by R.F. Christy and E. Segr\`e (while Fermi was absent) to
the scattering of neutrons and the albedo in the reflection of
neutrons, respectively.

The relevance of the present recovery is, therefore, clearly
twofold.

On a purely``scientific'' side, it is now evident that our
previous knowledge of the Fermi course was incomplete and, to some
extent (limited to the Halpern notes) misleading. As his usual,
Fermi was very accurate in the choice of the topics, that he
developed in detail and in a very clear manner, a peculiarity
which does not often emerge from the notes taken directly down in
class by students, and later arranged into the Halpern version.
Further studies on the original material present only in the
French notes are currently carrying out, and will probably reveal
some other precious results which will be of interest to
historians of physics (and to teachers of nuclear physics).

Instead we here dwell a bit more on the ``historical'' relevance
of the present recovery of the French notes.

As it is clear from what reported above, despite the
classification of the second part of the Fermi lectures by the
U.S. authorities, the detailed lectures on chain reactions and
pile physics (including applications to the bomb) were extensively
known to Chadwick, and thus to several British nuclear physicists,
as early as in 1947.

This is quite remarkable if compared with the persuasion of some
British people who believed that the U.S. authorities wanted a
postwar American nuclear monopoly. A decisive confirmation of this
view came, in fact, with the approval of the final version of the
McMahon Act, as already mentioned, which made illegal the passing
of ``restricted data'' to any foreign country, {\it including} the
United Kingdom. The seriousness of the consequences of the McMahon
Act for Anglo-American collaboration was readily recognized by
Chadwick \cite{BioMemo}, whose warning proved correct very soon,
since until 1948 no technical cooperation at all was set up and
until 1955 it was extremely limited. On the other hand, Chadwick
was a valiant supporter of a Britain's own nuclear project, mainly
directed towards the production in the United Kingdom of atomic
bombs and plants for producing fissile material, but also devoted
to possible medical applications. However, after the American
legislation on the control of nuclear energy was issued, British
scientists were (rightly) discouraged about the realization of a
Britain's own project without American technological cooperation,
since it appeared clear to them that key knowledge in this field
was then accumulated only in the United States. In this respect,
Chadwick's remark was again crucial: ``Are we so helpless,'' he
would ask, ``that we can do nothing without the U.S.?''
\cite{Gowing}.

The making of the French notes goes just in this direction and,
although no direct realization can at the moment be established,
in a sense it seems to support Chadwick's optimistic view. And
this is particularly intriguing, if his anxiety in avoiding
anything which might suggest to the Americans that the British
were trying to pry out secret information \cite{BioMemo} is taken
into account.

As a matter of fact, the British programme for nuclear energy,
initiated in the early postwar period, proved remarkably
successful. This was certainly due to the ability of British
scientists under the sensible guidance of Chadwick, but the French
notes of the Fermi course at Los Alamos proved to be at least very
useful in the effective launch of the British programme, and the
role played by Chadwick also to this regard was crucial.

Some previously obscure points on this matter are now seemingly
made clearer, and further studies even in this direction will led
to potentially interesting new results.


\subsection*{Acknowledgments}
The active and valuable cooperation of the staff of the
Chur\-chill Archive Centre, Cambridge (U.K.) and of the
Information Resource Centre of the U.S. Embassy in Rome is here
gratefully acknowledged. The author is also indebted to G. Miele
for his kind support and encouragement.

\end{document}